\documentclass[aps,prl,twocolumn,showpacs,superscriptaddress,groupedaddress,nofootinbib]{revtex4}
\usepackage[dvips]{graphicx}
\usepackage{epsfig}
\usepackage{subfigure}
\usepackage{float}
\usepackage{color}
\usepackage{ucs}
\usepackage[utf8x]{inputenc}
\usepackage{soul}
\usepackage{xr-hyper}

\begin{document}
\vspace*{-1in}
\title{A simple thermodynamic model for the hydrogen phase diagram}
\author{Ioan B Magdău, Miriam Marqu\'es, Balint Borgulya and Graeme J Ackland}
\affiliation{CSEC, SUPA, School of Physics and Astronomy, The University of Edinburgh, Edinburgh EH9 3JZ, United Kingdom}

\email{i.b.magdau@sms.ed.ac.uk, gjackland@ed.ac.uk}

\begin{abstract}
We describe a classical thermodynamic model that reproduces the main
features of the solid hydrogen phase diagram.  In particular, we show
how the general structure types that are found by electronic
structure calculations and the quantum nature of the protons can also
be understood from a classical viewpoint.  The model provides a
picture not only of crystal structure, but also for the anomalous
melting curve and insights into isotope effects, liquid metallisation
and InfraRed activity.  The existence of a classical picture for this
most quantum of condensed matter systems provides a surprising
extension of the correspondence principle of quantum mechanics, in
particular the equivalent effects of classical and quantum
uncertainty.
\end{abstract}
\maketitle


Solid hydrogen provides one of the greatest examples of complexity
emerging from a simple system. An equal mix of protons and electrons
is perhaps the most fundamental system in condensed matter.  Yet the
subtle interplay between thermodynamics and quantum mechanics produces
a phase diagram that has defied simple understanding.  The situation
has recently been further complicated by the discovery of a new phase
IV\cite{howie2012mixed,eremets2011conductive,zha2014raman}, reports of
further phases V and
VI\cite{dalladay2016evidence,eremets2016low,dias2016new}, and a
melting point maximum and
minimum\cite{Ashcroft2000,babaev2004superconductor,bonev2004quantum,howie2015raman,geng2015melt,chen2013quantum,mcmahon2012properties,goncharov2013hydrogen}.
Current theoretical work concentrates on finding candidate low energy
structures, characterized by symmetry, solving the electronic
structure alongside quantum protons across a range of temperature and
pressure. These computationally expensive numerical calculations
typically offer little insight into the underlying principles
determining the phase stability.  Here, instead of striving for
quantitative accuracy, we take the opposite approach, asking what is
the simplest atomic-level model that reproduces the qualitative phase
diagram. Our model is derived from studying energy-minimising
structures\cite{pickard2007structure,pickard2009structures,pickard2012density,geng2012high,liu2012quasi,labet2012freshall,monserrat2016hexagonal}
and trajectories of extensive molecular dynamics simulations performed
by us and
others\cite{liu2012room,magdau2013identification,magdau2013high,ackland2015appraisal}.
We identify three recurrent motifs from which we build a ``big
picture'' understanding of the thermodynamics of the phase diagram,
including metallization and isotope effects.

Currently, theoretical predictions of high pressure phases are based on
density functional calculations (DFT) using the PBE
functional.  Despite the deficiencies of this
method\cite{clay2014benchmarking,azadi2013fate}, improved methods
which include treatment of proton dynamics and electron correlation
lead to quantitative rather than qualitative changes to the calculated
phase
diagram\cite{azadi2013quantum,azadi2014dissociation,chen2014room,clay2014benchmarking,mcminis2015molecular,drummond2015quantum}.

The overall picture emerging from a combination of simulation,
spectroscopy and crystallography is as follows.  At low pressure Phase
I comprises quantum rotor molecules in a close packed structure.  At
very low temperature and increased pressure Phase II appears. Phase II
has X-ray diffraction very similar to Phase I, and is assumed to have
statically-ordered molecular orientations which minimise quadrupole
interactions\cite{mazin1997quantum,pickard2009structures}.  At higher
pressure Phase III is reported as a layered structure with weakly
bonded
molecules\cite{loubeyre2002optical,akahama2010evidence,pickard2012density,monserrat2016hexagonal}.
Phase IV, stable at higher temperatures, can be viewed as alternating
layers of Phase III-like weak molecules and Phase I-like strongly
bonded, rotating molecules. Phases named IV' and V, similar to IV, and
a premetallic phase VI have also been
reported\cite{dalladay2016evidence,eremets2016low,dias2016new}. 
The lowest known energy candidate for
phase II is $P2_1/c$ \cite{pickard2009structures} and for phase III
$P6_122$\cite{monserrat2016hexagonal} and $C2/c-24$ \cite{pickard2009structures}
below and above 200GPa
 respectively. The liquid, and
phases I and IV are calculated to have rotating molecules, leading to 
time-averaged symmetry higher than any static atomic arrangement.\cite{bonev2004quantum,liu2012room,magdau2013identification}.
The favored candidates for the metallic phase VI\cite{SilveraMetalarXiV}
are $Cmca$ and $I4/amd$\cite{johnson2000structure,pickard2007structure}

The melting curve 
has a strong positive slope at low pressures, but 
reaches a maximum at around 900K and 120GPa, and then drops.  The
Clapeyron slope flattens off once the solid transforms to
the denser phase
IV\cite{bonev2004quantum,morales2010evidence,howie2015raman,geng2015melt,chen2013quantum}.
The importance of quantum protons is highly
debated\cite{kitamura2000quantum,Ashcroft2000,morales2013nuclear}.  In
phase I the characteristic roton bands indicate that angular
momentum, $J$ is a good quantum number, and must combine with the nuclear
``para'' or ``ortho'' spin state to give an antisymmetric molecular wavefunction.
The  zero-point energy (ZPE), phonon free energy and associated
pressure can be approximated in
two ways, either via lattice dynamics and the quasiharmonic
approximation (LDQHA)\cite{bornhuang,ackland2002calculation}, and anharmonic corrections\cite{monserrat2013anharmonic} or via
path integral molecular dynamics (PIMD)\cite{marx1994ab}.  LDQHA
assumes delocalized, harmonic phonons, PIMD assumes distinguishable
atoms: neither approach describes freely rotating molecules.

In our model, free energy for each phase 
depends on its structure and its constituent
objects. 

The three objects in our model (named S, R and A) allow both quantum and
classical interpretation.  S is a spherical molecule, which
corresponds to a $J=0$ quantum rotor ground state, or a time-averaged
classical free rotor.  R is rodlike, corresponding
to the standard classical picture of two atoms connected by a covalent
bond, or the $J=1$ quantum rotor state. Finally, A 
represents simple spherical ``atoms'': these have unpaired
electrons which can explain electrical conductance within our model. For accounting purposes,
we consider pairs of type A ``atoms'', and de-dimensionalised units.

Only free energy {\it differences}
determine the phase diagram, so we can measure all energies, volumes and
entropies relative to an appropriate implicit reference which is
phase-independent, but without loss of generality may be pressure 
and temperature-dependent.

We set the covalent bond in both S and R objects to have energies
$U_S=U_R=-2$. Atoms are unbound, so U$_A$ is zero. The S objects have a
random orientation, which can be regarded as a classical entropy $S_S$.  These
values define the reduced (i.e. dimensionless) energy and entropy
units for the model.  Finally, we assign volumes to each object.
For the S molecule,
the volume $V_S$ represents the sphere swept out by the rotator, $V_R$
an ellipsoidal diatomic molecule, and $V_A$ a spherical atom, so clearly
$V_S>V_R>V_A$. The actual values used are given in Table
\ref{params}. In these reduced units, $V_S$ corresponds to a sphere of
radius 0.95, $V_R$ to a prolate ellipsoidal rod with the same major
axis and b/a=0.9, and $V_A$ is a sphere of radius 0.575.

\begin{table}[h]
\begin{flushleft}
\begin{tabular}{|l|ccc|}
\hline
{\bf type} $i$ &  S & R & A \\
energy $U_i$ &  -2 & -2 & 0 \\
entropy $S_i$ & 0.7 & 0.0 & 0.0 \\
volume $V_i$ & 3.6 & 2.7 & 0.8$\times$2 \\
\hline
\end{tabular}
\begin{tabular}{|l|ccccc|}
\hline
{\bf structure} $j$ &  I / VI & II & III & IV /V & liquid \\
packing c$_j$ &  0.74 & 0.71 & 0.74 & 0.79 & 0.71 \\
config. entropy S$_j$ & 0.0 & 0.0 & 0.0 & 0.0 & 0.75\\
bonding U$_j$ & 0.0 & -0.13 & 0.0 & 0.0 & 0.8 \\ 
\hline
\end{tabular}
\end{flushleft}
\caption{Parameters for objects and structures.  All volumes
  correspond to molecules (i.e. two atoms) and are described further in the SM.  
All values are strongly constrained by their well-defined physical meaning:
that the parameters all fall within reasonable bounds or can be
neglected entirely is a key result of the model. \label{params}
}
\end{table}

The model is formulated in terms of volumes, so to present the results on
a pressure-temperature phase diagram we require an equation of state.
We use
\[ x(P)= \frac{ (2P+1)}{(2P+0.15)},\label{eqv} \]
which describes a monotonic volume reduction by a  
factor of about 7 across the pressure range of interest.

The model's phases are as follows:

\begin{itemize}
\item 
Phase I has hexagonal
close packing (hcp) of S-objects.  $hcp$ is the most efficient packing of
spheres, with a packing fraction of $c_1=0.74$.
\[G_1=U_S+ x P V_S/c_1-xTS_S\]
\item Phase II, the ``broken symmetry'' phase, is
a structure in which molecules (R-objects) point in
directions to minimise the quadrupole-quadrupole interaction energy ($U_2$).
Packing is less efficient than Phase I, but the
overall density is higher because R-molecule have no rotation
(i.e. $V_R/c_2<V_S/c_1$).  
\[G_2=U_{2}+U_R+ xPV_R/c_2\]
\item 
Phase III is a more efficient packing of rods (R) than Phase II,
obtained at the cost of no longer
minimising the quadrupole interactions.
 \[G_3=U_R+ xPV_R/c_3 - xTS_R\]
\item  
Phase IV is  a mixed molecular-atomic layered
structure\cite{howie2012mixed} with molecular B-layers and atomic G-layers
respectively.    Our MD, showed such
structures with space group $P6/mmm$ as a time-average:
this comes from the B-layer molecules having
spherical symmetry 
and the time-averaged G-layer having sixfold symmetry.

We model the B-layers as composed of
S objects, and the G-layers as A objects. This 
``$SA_2$'' compound is equivalent to the MgB$_2$
structure, which is one of the most efficient packings of binary hard
spheres. 
Phase IV  incorporates all mixed phases IV, IV' and V\cite{dalladay2016evidence}.
 The subtle differences between these phases are not significant to this model, and are described later.
 \[G_4=G_5=(U_S+U_A)/2+ xP(V_S+V_A)/2c_4 - xT(S_S+S_A)/2\]
\item 
We treat the putative metallic Phase VI\cite{SilveraMetalarXiV} as a
close-packed atomic solid with type A objects.  In reality metallic
hydrogen may have a more open structure, but this is not yet known.
\[G_6= U_A +  xPV_A/c_1 - xTS_A\]
\item Liquid is a Boltzmann-weighted average of $S$, $R$ and $A$
  objects (labelled $i$), with additional configurational entropy $S_{liq}$ and 
 energy $U_{liq}$.
\[ G_{liq} = F_{liq} + PV_{liq}\]
with 
\[ F_{liq} = U_{liq}-TS_{liq} + \left(\sum_i ( U_i -TS_i)e^{-G_{i}/T}\right) / 
\mathcal{Z} \]
\[V_{liq} = \frac{x}{C_{liq}\mathcal{Z}}\left(\sum_i V_i\exp(-G_{i}/T)
\right)\] 
where $i$ indicates sums over S, R and A.
\[G_{i}= U_i + xPV_i - TS_i;\hspace{1cm}\mathcal{Z} =  \sum_i\exp(-G_{i}/T)
\]
\end{itemize}

Terms set to zero in Table.\ref{params} are ignored. 

For the structural contributions to free energy, we assume that the
only significant deviation from intermolecular bonding between
different phases at the same (P,T) conditions comes from quadrupole
alignment in phase II, and that the liquid has higher
configurational entropy and reduced cohesive energy.
We ignore energy and entropy contributions that are similar
for all structures:  these give a structure-independent contribution
to the free energy which does not affect the relative free energies. 
which determine the
phases diagram.

The final parameter describes zero point vibration.  LDQHA and PIMD
calculations have shown that ZPE is the dominant contribution from
nuclear quantum effects, and the effect on the phase diagram is, to a
first approximation, a shift of all phase boundaries to lower
temperatures\cite{Morales2013towards}.  We understand this as a loose
equivalence of quantum and thermal oscillations, and account for it by
shifting the T=0 axis up by 0.45.  This shift is the only
isotope-dependent effect in the model, it distinguishes hydrogen from
deuterium, for which it is smaller.  The I-II phase boundary in
deuterium is then at lower temperatures than for hydrogen, and cuts
the T=0 axis at lower pressure, as observed.

\begin{figure}[H]
\includegraphics[width=85mm]{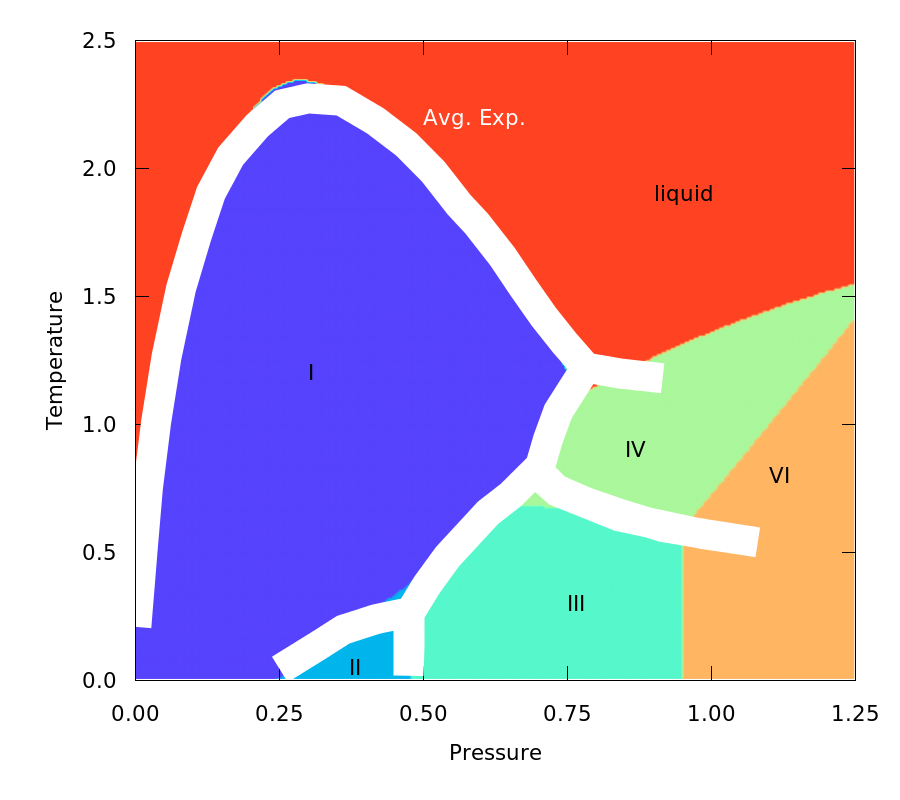}
\caption{Phase diagram. Colors depict the
  phase with lowest Gibbs free energy. Grey lines show the current
  experimental situation, with approximate uncertainty, (see Supplemental Material.  Phase IV'
  and V are considered as continuous
  with Phase IV.  Temperature and Pressure are given here in the
  reduced units of the model, for comparison to experimental GPa
  and K units, pressure should be scaled by 240 and temperature by
  370.
\label{simplePD}}
\end{figure}

Remarkably, the phase diagram produced 
(Fig. \ref{simplePD}) for any sensible choice of parameters has
stability regions 
for the six phases in the correct regions of PT space  
and 
a melting curve with a maximum.

This gives some insights into the nature of the various phases.  The
melting temperature maximum means that the liquid has a higher
compressibility than the solid.  In our model this is because the
large S objects in the liquid increasingly convert to smaller R and A
objects with pressure.  The competing phase I has only large S
objects, so becomes less favoured at pressure, despite its
close-packing.  Phase IV is assumed denser than the liquid, so its
melting point increases with pressure.

The model suggests a novel interpretation of the liquid
insulator/metal transition\cite{knudson2015direct,ackland2015bearing}.
Assuming that molecules (S,R) have localised electrons and atoms
have delocalised electrons, conduction occurs once there are
sufficient complete paths via neighbouring A objects for electron
hopping to percolate: this can occur either at high temperature, where
all objects are equally likely, or at high pressure where the fraction
of smaller A objects is increased.

Phase IV has a free energy advantage over
the purely atomic phase thanks to its molecule bonding, and over the pure
molecular phase I because of its efficient packing of molecules and
atoms.  It is stablized against phase III by the entropy of the
rotating S molecules.

The phase diagram shows a positive Clapeyron
slope between the atomic (metallic) phase VI and the semiconducting
phase IV.  There is no thermodynamic reason why a material cannot 
become metallic on cooling, but it is very unusual.
Here, it occurs because of the extra rotor entropy  $S_S$, 
compared with the zero value of $S_A$.

The model does not include a zero-temperature quantum liquid phase at
high-T.  This is mainly because we choose not to make the ZPE offset
pressure dependent. It is possible to choose parameters for
which the melting point goes to zero at high pressure.

\begin{figure}[H]
\includegraphics[width=\linewidth]{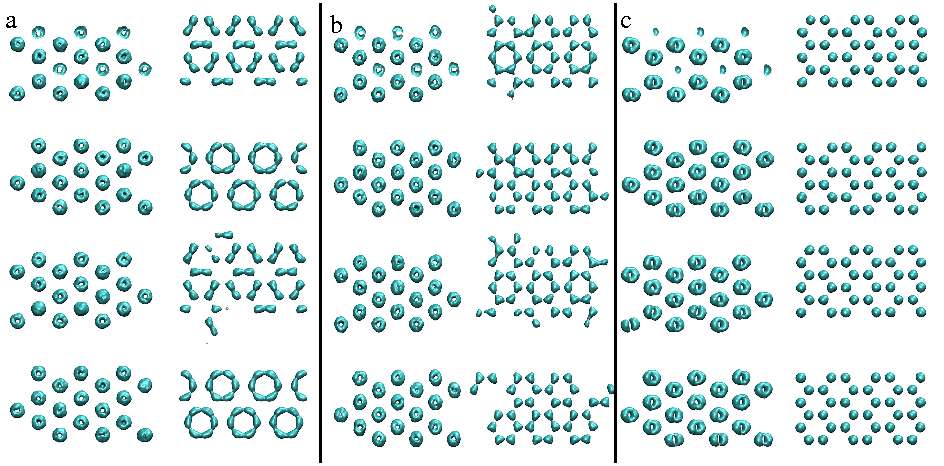}
\caption{Isosurfaces of time-averaged probability density for
  indistinguishable atoms from 8-layer AIMD
  simulations at different pressures. (a) 250GPa G'BG''B ``Phase IV''
  (b) 325GPa BG' ``Phase V'' (c) 400GPa BG ``atomic-molecular''; 
\label{MDresultsISO}}
\end{figure}

Perhaps the most serious simplification entailed by the model compared
with our ab initio MD\cite{Datashare} comes in the treatment of the
so-called graphene-like G-layers of phase IV.  
The structure of phase IV seems well described by ab initio molecular
dynamics, but although a new Phase V
was reported earlier this year, our extensive ab initio molecular
dynamics calculations in this pressure/temperature regime show 
changes in the dynamics, rather than in the time-averaged structure. 
Currently, phases IV and V  are treated the same 
in our model, as mixed atomic-molecular structures.  In MD
simulations\cite{Datashare,liu2012room,magdau2013identification,youtube}
the G-layer atoms are observed to pair up into short-lived, weakly bound
molecules (Fig. \ref{MDresultsISO}). 
We introduced new analysis methods to
monitor bond breaking and reconstruction in DFT-MD calculations.
This showed that the
MgB$_2$ structure is reasonable as a long time-average, 
but there are subtle changes in symmetry with pressure.

The MD implies that the G-layer can be described by decoration of a
hexagonal lattice, and the subtle experimental differences between
Phases IV, IV' and V are also consistent with this.  Figure
\ref{Glayer} gives a schematic view of three possible decorations.  In
MgB$_2$, the atoms would be located on the vertices of the lattice
(labelled G$^a$), and molecular dynamics at high pressure shows this
structure on average.  However, at lower pressures the atoms pair up
to form weakly-bonded molecules, the weakness evidenced by low
frequency vibrons. The structure has a four-layer BG'BG'' repeat: in
the G'' arrangement the molecules form
trimers\cite{lesar1981likelihood} with six atoms inside one in three
of the ``cells'' of the honeycomb network. In MD, the trimer rotates
as a unit.  In the G' arrangement, the molecules are located on the
boundaries between cells.  

In static relaxation, the B-layer molecules cannot have hexagonal
symmetry, and this symmetry-breaking induces further symmetry breaking
in the G-layer.  Structure searches have revealed a panoply of such phases
\cite{pickard2007structure,geng2012high,liu2012quasi}

MD shows continuous transitions between G-layer decorations
(Fig. \ref{MDresultsISO}).  At the onset of phase IV, we find a four
layer stacking with alternating BG'BG'' layers.  The yellow-centered
atoms and gray rhombus in Fig 3 show the elegance of this arrangement:
notice how the G'' trimer is located above the cell in G' which has no
molecules on its boundaries.  As pressure increases, all G-layers
adopt the G' arrangement at the long time scale, whereas at the short
time scales, trimers rebond faster and faster: this is our description
for ``phase V'' .  At still higher pressures the atomic G$^a$-layers
are observed.

Phase III has previously been reported as a ``layered'' structure, but
the logic here requires it to be efficiently packed.  This is counter
to current understanding\footnote{Unnamed referee during review
  process}, and we have carried out further DFT calculations of the
two most likely candidates. Whereas previous work has focussed on
atoms, in figure \ref{miriam} we show that the ELF
isosurfaces of the H$_2$ molecules are close to ellipsoidal, and the
molecule centres themselves are arranged very close to hcp.  

This is represented by ordered R-objects in our model.  The
fundamental description of phase III is close-packing of molecules.
Candidate structures for
phase III are based on layers like that shown in figure \ref{miriam},
with molecules pointing in one of three possible directions.  
The next layer fits
efficiently with 2/3 molecules located above the larger interstices
and the third above triple-triangular interstice in the centre of the
figure.  The orientation of the molecules is of secondary importance, but it is
this which defines the crystal symmetry. All
near-neighbour molecules in a layer have different orientations.  The $C2/c-12$
structure has a two-layer repeat stacking, with molecules two hcp
layers above pointing in the same direction.  $C2/c-24$ has a
four-layer repeat stacking, while the lowest energy $P6_122$ structure
has a six layer repeat, cycling through all three possible
orientations and giving it the highest symmetry.

It can also be seen that to maintain efficient packing the molecules
become asymmetric: the midpoint between nuclei is not precisely at the
centre of the electron distribution, nor on the hcp lattice site.
This causes the molecule to obtain a dipole moment, which is in turn
responsible for the strong IR signal which characterises Phase III.

In all these candidate phases, the rods lie in the plane, so according
to the model the c/a ratio should be less than ideal ($\sqrt{8/3}$ for
a two-layer repeat). DFT calculation for $P6_122$, for which c/a is
uniquely defined, gives a value of 1.549 at 150GPa dropping to 1.541
at 350GPa.  

\begin{figure}[H]
\includegraphics[width=\linewidth]{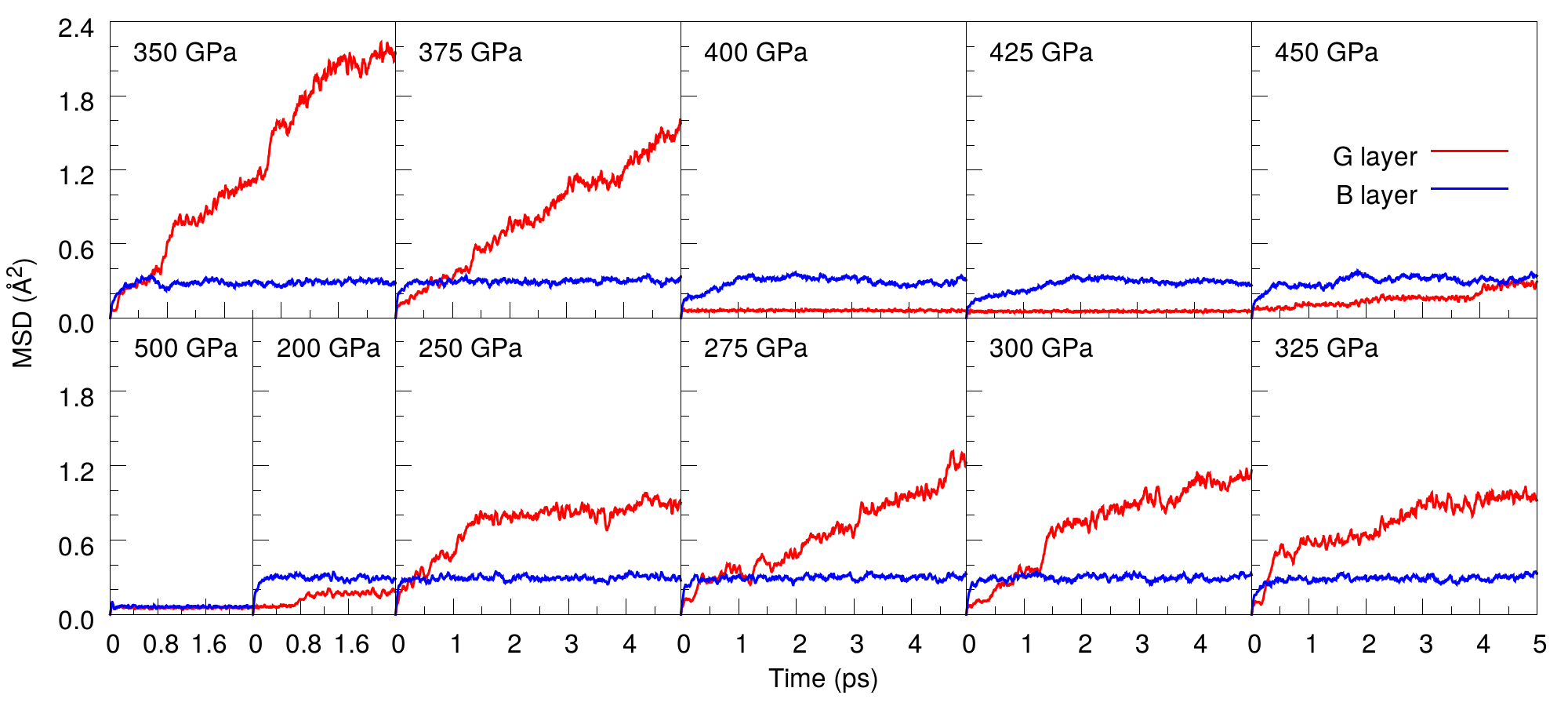}
\caption{ Mean squared displacements at 300K from ab initio MD
  simulations at various pressures.  Note the saturation of MSD for
  rotating molecules in B-layers, larger saturated G-layer MSD for
  rotating trimers in BG''BG' (250-325GPa), linearly increasing
  diffusive MSD for BG' (350-375GPa) and no diffusion for atomic
  G-layers (400-450GPa).  This different dynamic behavior
  distinguishes Phases IV, IV' and V in the MD and in spectrocopy, but
  it is debatable whether they are thermodynamically distinct phases,
  so they are all treated equivalently in the model.}
\label{MDresultsMSD}
\end{figure}

Figure \ref{MDresultsMSD} shows how the diffusion of phase IV varies with
pressure and Figure \ref{Glayer} gives an insight into the process of
the diffusion.
\begin{itemize}
\item 
In G'' layers it is possible for two correlated events to occur in the
trimers:
bond breaking where the definition of
molecules changes between two permutations;
and trimer rotation through 60$^o$.
These two processes are distinct in the classical  MD,
but equivalent for indistinguishable quantum protons. In
either case, all atoms remain within the same hexagonal cell and no
diffusion is possible. This rebonding leads to short
lifetime of molecular vibration in the G-layer, and consequent
broadening of the Raman vibron in addition to anharmoic effects\cite{monserrat2013anharmonic}.

\item In G$^a$ layers, 
diffusion cannot occur, except via vacancies.

\item In G' layers molecules are located between two cells.  A trimer
rotation through 120$^o$ leaves the pattern unchanged, however a sequence of
such rotations in neighbouring cells can move the molecule through the
lattice, giving rise to true diffusion.  In the BG''BG' stacking such
rotation is suppressed because the G'' hexagons impose 
ordering in the G' layer.
\end{itemize}

In MD we find that diffusion
in the
BG'BG'' and BG$^a$ structures,
is low, but for the BG' structure it is significant.
This additional diffusion implies increased broadening of
spectroscopic lines with increasing pressure - the most notable
signature of Phase V.

To summarize, we have built a model for the hydrogen phase
diagram based around simple concepts and a few descriptive parameters.
The model is robust: any sensible choice for the parameters 
gives a phase diagram including the known phases and unusual behaviour of
the melting curve.  While there is no doubt that a quantitative
theoretical description of the phase diagram requires complex quantum
treatment of both protons and electron, it is remarkable that the
overall picture can be captured with classical free energies.

In addition to reproducing known phases, the model makes a number of
predictions which can be used to guide analysis of future, more detailed calculations, namely 

a) The melting point maximum is due to the liquid being a mix of large
and small objects.

b) The liquid metal-insulator transition has a percolation/localization character

c) Phase III should be thought of as closely packed molecules,
somewhat elongated but close to spherical, rather than layers of
atoms.

d) Isotope effects are generally reported at lower pressure in
deuterium compared to hydrogen: this could equivalently be described
as shifted to higher temperature, which is our approach.  The
consequence is that isotope effects are far more pronounced in 
transitions with shallow Clapeyron slopes.

e) Efficient packing of ellipsoids in Phase III leads to molecular
asymmetry, a dipole moment, and explains the strong IR signal.

f) The metallic Phase VI of our model need not be closely packed, the increased density arises from the atoms being smaller than molecules.

g) Phase IV adopts a time-averaged structure which represents the
known most efficient close-packing of binary hard spheres.  Hence it
is stabilized by packing effects as well as entropy.

\begin{acknowledgments}We thank E.Gregoryanz, A.Hermann, M.Marques,
  C.Pickard, I. Silvera, B.Monserrat and M.Martinez-Canales for many
  useful discussions.  We thank EPSRC for computing time (UKCP grant
  K01465X) and for a studentship (IBM) GJA was supported by an ERC
  fellowship ``Hecate'' and a Royal Society Wolfson fellowship.
\end{acknowledgments}

\onecolumngrid

\begin{figure}[H]
\includegraphics[width=0.45\linewidth]{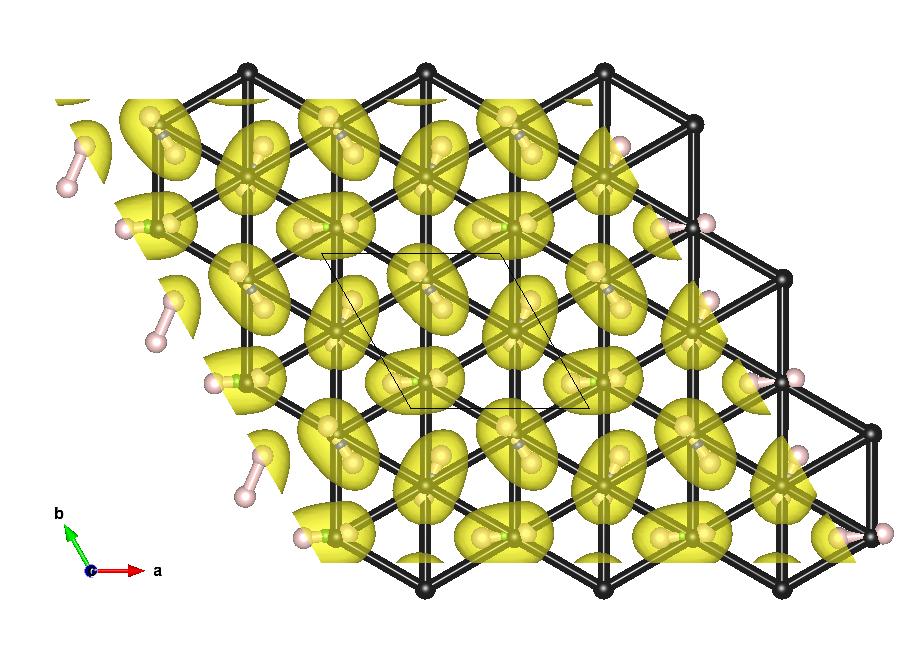}
\includegraphics[width=0.45\linewidth]{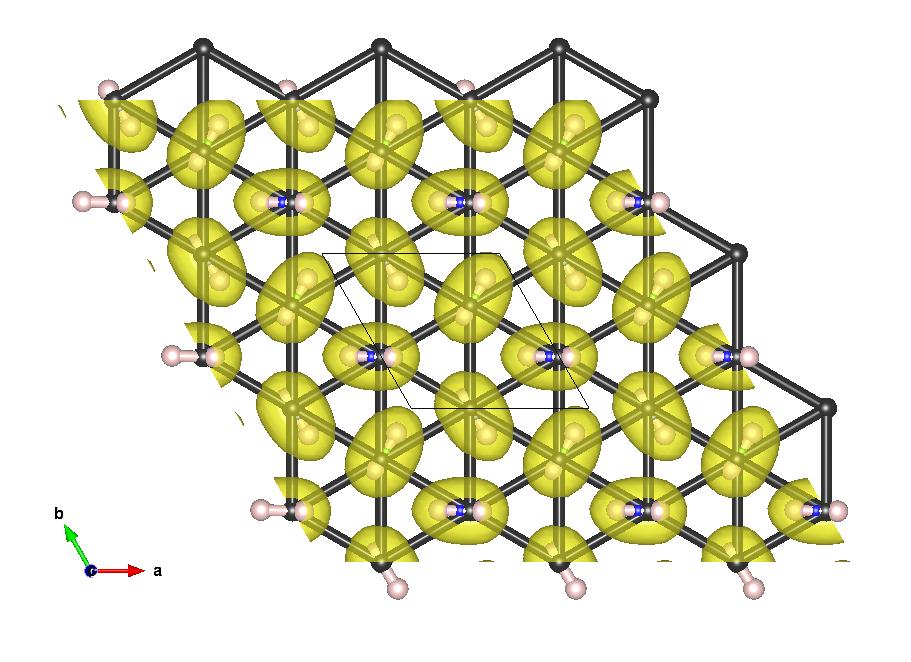}
\caption{
Electron Localization Function isosurface (ELF=0.5) for one plane of phase III candidate
structures (a) $C2/c$  \cite{pickard2009structures}  and (b) $P6_122$ \cite{monserrat2016hexagonal}, illustrating the rationale for 
  modelling it in terms of
efficient packing of rod-like molecules. Pink spheres correspond to the hydrogen atoms, whereas smaller blue (green) spheres are located at the midpoints of the two types of molecules, with slightly larger (shorter) bond lengths.
Black spheres and lines represent the hcp packing, and show that the
molecular centers can be regarded as  almost close-packed.}
\label{miriam}
\end{figure}

\newpage 

\twocolumngrid

\begin{figure}[H]
\includegraphics[width=85mm]{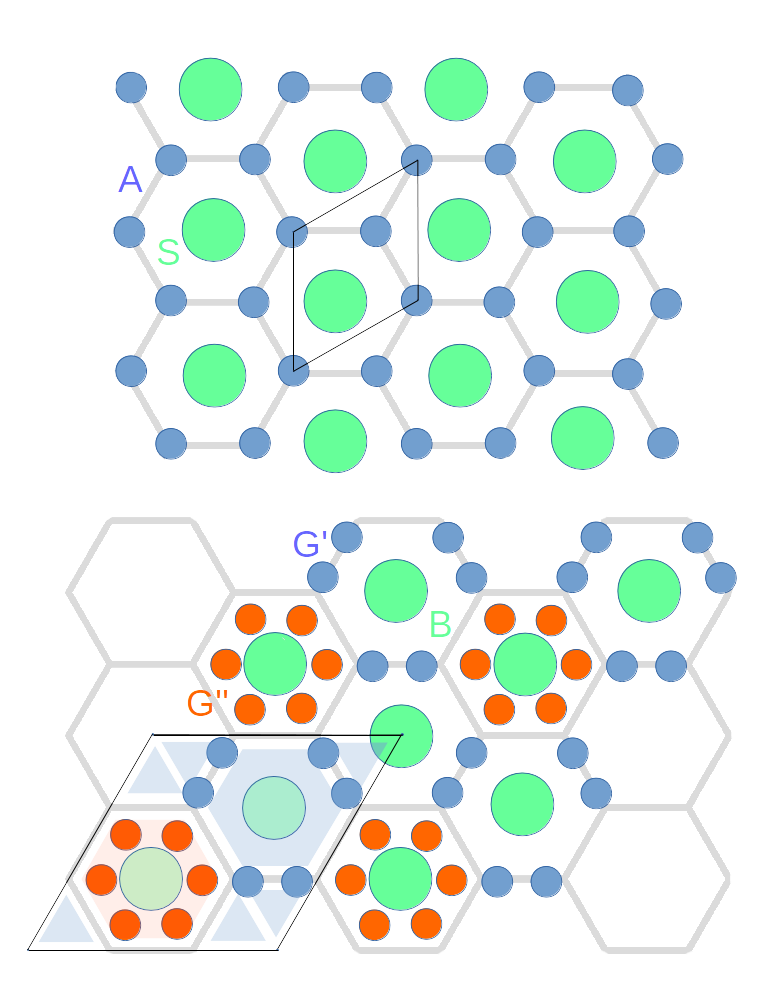}
\caption{Idealised geometric stacking patterns for phases IV/IV'/V as
  deduced from Fig.\ref{MDresultsISO}: (upper) Two-layer $P6/mmm$
MgB$_2$ structure, with S objects on the Mg site and A on the B
sites. Thick grey lines showing hexagonal symmetry, thin black line
showing primitive cell.  Note that the $P6/mmm$ requires only that the
S-molecule rotates about the z-axis, appearing as a donut in Fig.\ref{MDresultsISO}. (lower) Four-layer BG'BG'' broken
symmetry structure with weakly-bonded R-type molecules: blue: G', red
G'', thin black line indicates unit cell of BG'BG'' stacking
structure, with G-layers at different heights.}
\label{Glayer}
\end{figure}

\begin{figure}[H]
\includegraphics[width=85mm]{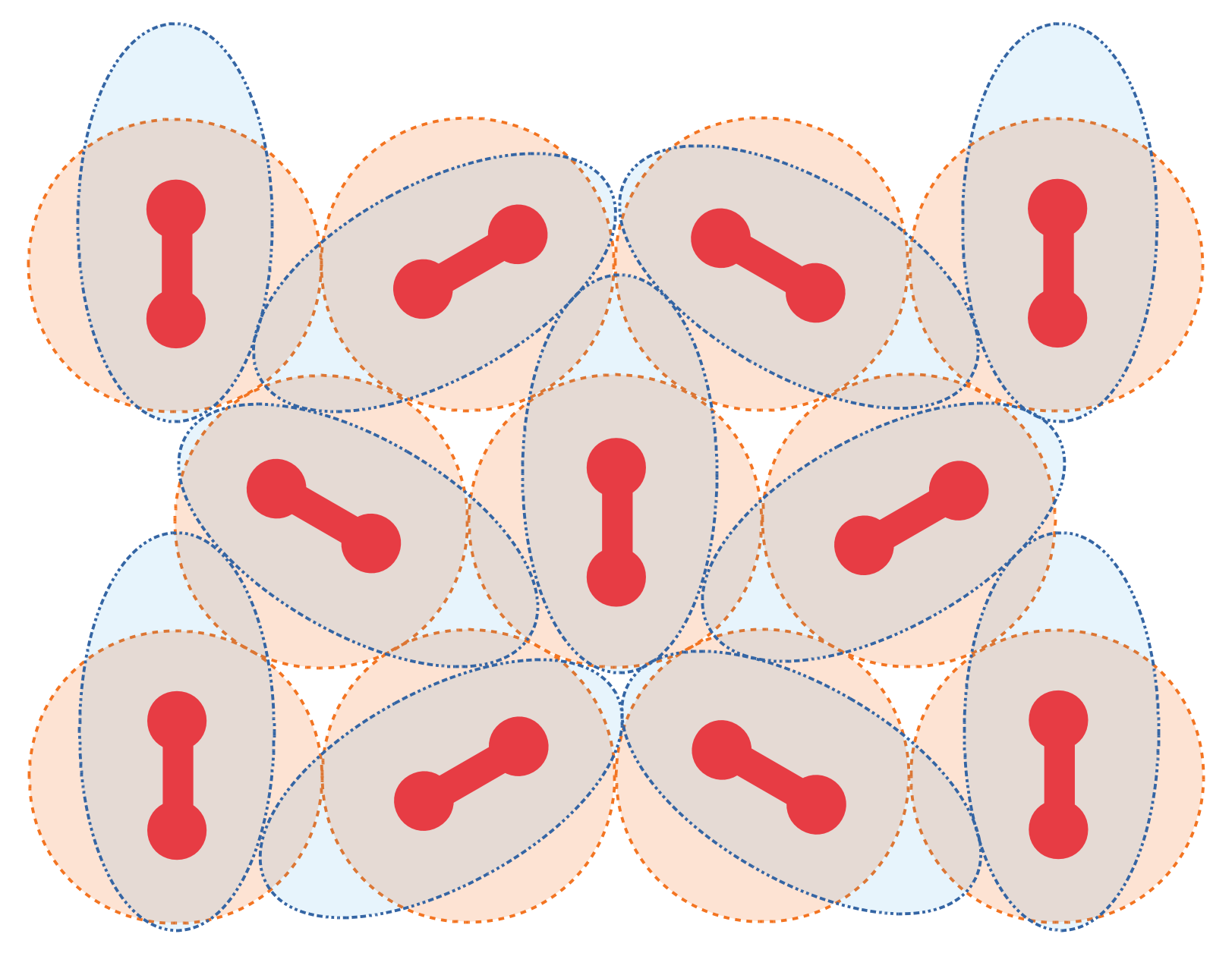}
\caption{Geometric stacking patterns for Phase III, where
  conventional ``layered'' structure molecules (red dumbbells) are
  centred on spheres, illustrating the ``hexagonal close packing''
  interpretation.  The dumbbell orientation is common to the proposed
  $C2c$ or $P6_122$\cite{monserrat2016hexagonal}.  The ellipses show
  how the orientation of R objects gives efficient packing of
  ellipsoids as a distortion from hcp\cite{akahama2010evidence}. 
\label{miriam}}
\end{figure}

\bibliographystyle{apsrev}
\bibliography{Refs}

\end{document}